\documentstyle[prl,epsf,aps]{revtex}

\normalsize

\begin{document}

\title{Effect of differences in proton and neutron density
             distributions on fission barriers}

\author{              J.F. Berger\\
  {\it Commissariat … l'Energie Atomique, Service de Physique Nucl'aire}\\
  {\it  B.P. 12, 91680 BruyŠres-le-Chƒtel, France }\\
                      K. Pomorski\\
 {\it Theoretical Physics Department, University M.C.S., Lublin, Poland}}

\date{\today}

\maketitle

\begin{abstract}
The neutron and proton density distributions obtained in constrained
Hartree--Fock--Bogolyubov calculations with the Gogny force along the
fission
paths of $^{232}$Th, $^{236}$U, $^{238}$U and $^{240}$Pu are analyzed.
  Significant differences in the multipole deformations of neutron and
proton densities are found.
The effect on potential energy surfaces and on barrier heights of an
additional constraint imposing similar spatial distributions to neutrons
and protons, as assumed in macroscopic--microscopic models, is studied.


\end{abstract}
\bigskip


Experimental analyses of, e.g. electron and $\alpha$--particle
scattering, pionic atoms, anihilation of antiprotons show that, in most
nuclei, neutrons and protons have close but different r.m.s. radii
\cite{Ba89}.
The main reasons for this difference are unequal numbers of neutrons and
protons, and  the Coulomb interaction between protons.

  In the last twenty-five years, fully microscopic approaches employing
parameterized forms of the effective interaction between nucleons within a
formalism of the Hartree-Fock type have been developed, which usually
reproduce experimental proton and neutron r.m.s. radii and their difference
in a satisfactory way \cite{Go75,Qu78,DG80,Ba95,Po97,Wa98,Pat99}.
Corrections coming from oscillations of the mean-field in principle
have to be added, which slightly increase nucleon radii. However, they
often may be neglected, except in a few closed shell nuclei and in
nuclei exhibiting shape coexistence \cite{Go79,Gi76,Gi82}.

  Experimental r.m.s. charge radii and isotopic shifts are also successfully
reproduced by macroscopic--microscopic methods based on the liquid drop model
and the Strutinsky shell correction technique \cite{Ne93,Ne95,Lo95}.
  Contrary to microscopic approaches, the proton and neutron mean-fields
are not self-consistently determined, but taken as
local potentials of prescribed forms -- Nilsson or Saxon--Woods --, with
identical spatial deformations.

  In deformed nuclei, neutron and proton nuclear density distributions are
expected to have not only different radii, but also different shapes, i.e.
different quadrupole and higher multipole deformations. In Ref.
\cite{Po97},
an analysis of theoretical densities based on a surface multipole moment
expansion showed that significant differences between neutron and proton
deformations often occur.

  The aim of the present investigation is to check how large the
differences between neutron and proton multipole deformations along
paths to fission are, and to what extent fission barriers depend on
such differences.
  This study is based on the constrained Hartree--Fock--Bogolyubov (HFB)
approach with the Gogny effective interaction which has been extensively
used
several years ago to describe actinide fission \cite{Be89}.
Four well-known actinide nuclei,
$^{232}$Th, $^{236}$U, $^{238}$U and $^{240}$Pu are considered.

  Moments of the multipole components of the neutron and proton densities are
first calculated as functions of quadrupole deformation using the method
described in the next Section.
  In a second step, a constraint is introduced in the self-consistent
calculation in order to impose identical spatial distributions to
neutrons and protons.
  The resulting effect on the shape and height of fission barriers can be
viewed as an estimate of the influence of the assumption usually made in
macroscopic-microscopic Strutinsky type models (see e.g.
\cite{St89,Sob,Mol}),
that the neutron and proton potentials have the same multipole
deformations.


  The constrained HFB calculations have been performed following the method
described in Ref.\cite{Be89}. Axial symmetry of the nuclear shapes has been
assumed. Nuclear deformation along fission paths has been generated
by means of a linear constraint on the nuclear mass quadrupole moment
$\widehat{Q}_{20} = \sum_{i=1}^{A} r_{i}^{2} P_{2}(\cos \theta_{i})$.
The two-body effective nucleon-nucleon
interaction has been taken in the form proposed by Gogny \cite{DG80},
with the set of parameters D1S \cite{Be91} adopted since 1983. Let us
recall
that this finite range effective force has been proved to give a very
satisfactory {\it ab initio} description of
the average field and pairing correlations in nuclei, and also of
actinide fission barriers.

  Generalized multipole moments $Q^k_\lambda$ of the self-consistent
density
distributions $\rho(\vec r)$ are defined by the following integrals:
\begin{equation}
 Q^k_\lambda = \int r^k P_\lambda(\cos\theta) \, \rho(\vec r) \, d^3\vec r
\end{equation}
where $P_\lambda$ is a Legendre polynomial of order $\lambda$.
The average $\beta$--deformation of multipole $\lambda$ can be taken
proportional to
the ratio of the moment $Q^k_\lambda$ to the monopole moment $Q^k_0$:
\begin{equation}
 \beta_\lambda(k) = { \sqrt{4\pi (2\lambda+1)} \over k+3}{Q^k_\lambda\over
  Q^k_0}\,. \end{equation}
This definition yields deformation parameters close to the
$\beta_\lambda$ commonly used to define the shape of the nuclear surface
in the liquid drop model :
\begin{equation}
R(\theta,\phi) = R_0(\{\beta_\lambda\}) \left( 1 + \sum_{\lambda=1}^m
\beta_\lambda
                 Y_{\lambda 0}(\theta,\phi)\right) \,\,.
\end{equation}
In particular, for small deviations from the sphere
and for a uniform density distribution~:
\begin{equation}
 \beta_\lambda(k) = \beta_\lambda ,
\end{equation}
for any value of $k$ in Eq. (1).

  In Fig. 1, the fission barriers of $^{232}$Th, $^{236}$U, $^{238}$U, and
$^{240}$Pu are plotted as functions of the mean value of the
nucleus mass quadrupole moment $\langle \widehat{Q}_{20} \rangle$.
The solid line corresponds to left-right reflection symmetric
nuclear shapes, while the dashed line represents the more general asymmetric
case. As expected, fission barriers including asymmetric shapes
are lower in these nuclei beyond the isomeric minimum.
HFB energies along the fission paths have been corrected in the usual
way \cite{Gi79} from the spurious zero-point energies associated with the
fluctuations of the center of mass position, of the angular orientation
and of the quadrupole deformation of the HFB states.
Both one-body and two-body center of mass corrections have been computed.
  Let us mention that the lowering of the first hump of barriers due to
triaxial instability is not included in the curves of Fig. 1.

The multipole deformation parameters: quadrupole $\beta_2$,
octupole $\beta_3$ and hexadecapole $\beta_4$ defined by Eq. (2) have been
evaluated along the different fission barriers of Fig. 1.
As an example, the multipole deformations of the total nuclear density
along the symmetric and asymmetric barriers of
$^{232}$Th are plotted in the upper-left part of Fig. 2.
In this analysis the surface multipole moment, i.e. $k = 2$, has been
chosen.
Deformations obtained with $k = 4$ are not shown
as they are close to those computed with $k = 2$.

Inserting neutron (resp. proton) density distributions into formula (1)
allows one to get
information about the neutron (resp. proton) multipole deformations.
The other three diagrams in Fig. 2. display the
differences $\beta^n_\lambda - \beta^p_\lambda$
between neutron and proton deformations in $^{232}$Th.
One observes that they strongly depend on deformation and
are always negative for prolate deformations. Large values
are obtained beyond the second minimum of the fission barrier,
especially for octupole and hexadecapole deformations.
There, the relative neutron-proton deformation difference
$\vert \beta^n_\lambda - \beta^p_\lambda \vert / \beta_\lambda $
reaches 4\% in the quadrupole case, and exceeds 10\%
for $\lambda = 3$ and $4$.

The above neutron-proton differences exhibit a similar behavior
in $^{236}$U, $^{238}$U and $^{240}$Pu, although their
magnitudes are not as large as in $^{232}$Th. As in the latter nucleus,
the largest numbers are obtained for octupole and hexadecapole
deformations.
  From these results one can conclude that the rearrangement of the nuclear
structure along fission paths leads to an increase of the difference
between
neutron and proton deformations, the proton multipole deformations becoming
significantly larger than neutron ones beyond the fission isomer
potential minimum.

In order to estimate the influence on fission barriers of the differences
found between the neutron and proton spatial distributions,
the HFB calculations have been performed again, adding
a new constraint ensuring that all multipole deformations
$\beta_\lambda(n)$ of protons and neutrons are equal.
This has been done by imposing, at each iteration of the HFB procedure,
that
the neutron and proton density matrices are proportional to each other:
\begin{equation}
 \rho_n = {N\over Z} \rho_p
\end{equation}
No such condition has been imposed on the neutron and proton pairing
tensors.

 As expected within a variational framework, the HFB energies computed with
the additional constraint are found higher than those computed without
this subsidiary condition.
The differences  $\delta E_{\rm den}$  between the
HFB energies computed with and without the additional constraint along the
fission paths of $^{232}$Th, $^{236-238}$U and $^{240}$Pu are plotted in
Fig. 3.
As in Figs. 1 and 2, the solid lines represent the fission paths
for reflection symmetric nuclear shapes, while the dashed lines correspond to
reflection asymmetric ones. One can see that the  $\delta E_{\rm den}$
differences are approximately 1.5~MeV on the average in
$^{240}$Pu and $^{236-238}$U, with fluctuations having an amplitude of the
order of 1~MeV.
In $^{232}$Th, the average energy difference is slightly smaller, while
stronger oscillations can be observed.

As a consequence, imposing condition (5) leads, in some cases, to an
increase of fission barrier heights of $\sim$1~MeV. This can be seen
in Fig. 4, where the fission barriers computed with (dashed line)
and without (solid line) constraint (5) are plotted.
Here, only the barriers corresponding to reflection asymmetric shapes are
displayed. In order to better compare the two calculations,
the dashed curves are shifted downward in order that the ground
state minima in the dashed and solid curves coincide.
  One observes that, in spite of this shift, constraint (5) almost always
leads to an increase of the heights of the barrier maxima and of the
isomeric secondary minima.
  The biggest change occurs in $^{238}$U, where the height of the first
barrier increases by 1 MeV.
A smaller but sizable ($\simeq$~.4~--~.7~MeV) increase of first barrier heights
of the other three nuclei can be seen.
In all nuclei except $^{240}$Pu, similar increases are obtained for second
barriers and isomer well energies.
In addition, the shape of the second barrier is slightly altered
in $^{232}$Th and, to a lesser extent, in the other three nuclei.

  In order to better illustrate the consequences of these barrier increases,
their influence on spontaneous fission lifetimes has been estimated.
  Barrier penetration factors have been calculated using the WKB
approximation, with collective inertia parameters computed from the method
given in Ref.\cite{Ran76}. As a result,
constaint (5) leads to spontaneous fission lifetimes
increased by factors 9.8, 6.6, 22.4 and 2 in $^{232}$Th, $^{236}$U, $^{238}$U
and $^{240}$Pu, respectively.


This work shows that the multipole deformations of the proton
and neutron density distributions of fissioning nuclei are far from being equal.
The relative difference between them often exceeds 10\% and
undergoes large variations.
This means that the thickness of the neutron skin
does not remain constant as the fissioning nucleus elongates.
The effect on the nuclear binding energy of these deformation differences
is found to be approximately
1.5 MeV, with fluctuations of the order of 1 MeV.
Clearly, these numbers are not negligible compared to typical fission
barrier heights.

  As mentioned earlier, macroscopic-microscopic calculations of
potential energy surfaces of fissioning nuclei assume equal deformations of
protons and neutron distributions.
  In view of the above results, one is tempted to infer that these
calculations predict fission barriers that are systematically
$\simeq$ 1 MeV too high.
  One must note that only the fluctuating part of $\delta E_{\rm
den}$ can actually influence barrier heights calculated with the
macroscopic-microscopic approach, since the average value of this
difference
can be taken into account in the fitting of the parameters of the
macroscopic (e.g. liquid drop) model.
  Still, as shown above, neglecting neutron-proton deformation differences
can induce an overestimation of spontaneous fission lifetimes by several
orders of magnitude in heavy and superheavy nuclei.

  This result suggests that it might be useful to generalize the
currently used macroscopic-microscopic approaches in order to
allow protons and neutrons to have different multipole deformations.
One could start from the energy prescription :
\begin{equation}
 E_{\rm Strut}(\{\beta^p_\lambda\}, \{\beta^n_\lambda\}) =
 E_{\rm macr}(\{\beta^p_\lambda\}, \{\beta^n_\lambda\}) +
\delta E^p_{\rm micr} (\{\beta^p_\lambda\}) + \delta E^n_{\rm
micr} (\{\beta^n_\lambda\})
\end{equation}
where both the macroscopic and and microscopic components depend on proton
and neutron deformations. Then, performing a minimization with respect to
e.g., the neutron multipole deformations while keeping proton deformations
as independent variables, should give a more realistic description of
potential energy surfaces. Details of such a method as, for instance, the
way to generalize macroscopic models in order to include
different deformations for protons and neutrons, are left for future study.
\medskip

\noindent
This work is partially financed by the Polish Committee of Scientific
Research under Contract No. 2~P03B 011 12.

\newpage

\centerline{\large {\bf Figure captions}}

\bigskip
\begin{enumerate}
\item Fission barriers of $^{232}$Th, $^{236}$U, $^{238}$U and $^{240}$Pu
obtained with the HFB method and the Gogny effective interaction. They are
drawn as functions of the nucleus total (mass) quadrupole moments.
Center of mass, rotational and vibrational corrections are included, as
explained in the text.
  The solid lines represent the fission barriers for reflection symmetric
shapes, while the dashed lines correspond to asymmetric fission.

\item Multipole deformations $\beta_\lambda , \,\,\lambda=$2,3,4 of the
$^{232}$Th density (upper left), and the differences $\beta^n_\lambda
- \beta^p_\lambda$ between neutron and proton multipole deformations
for symmetric (solid curves) and asymmetric (dashed curves)
fission of $^{232}$Th, as  functions of the total quadrupole moment.

\item Differences between the fission barriers obtained with and without
constraint (5) imposing equal deformations to neutrons and protons,
in $^{232}$Th, $^{236}$U, $^{238}$U and $^{240}$Pu.

\item Fission barriers for the asymmetric fission of
$^{232}$Th, $^{236}$U, $^{238}$U and $^{240}$Pu,
obtained with  (dashed curves) and without (solid curves)
constraint (5) imposing equal deformations to neutrons and protons.
The dashed curves are shifted downward in order that ground state
minima coincide.

\end{enumerate}

\pagestyle{empty}
\newpage
\begin{figure}
\epsfxsize=160mm \epsfbox{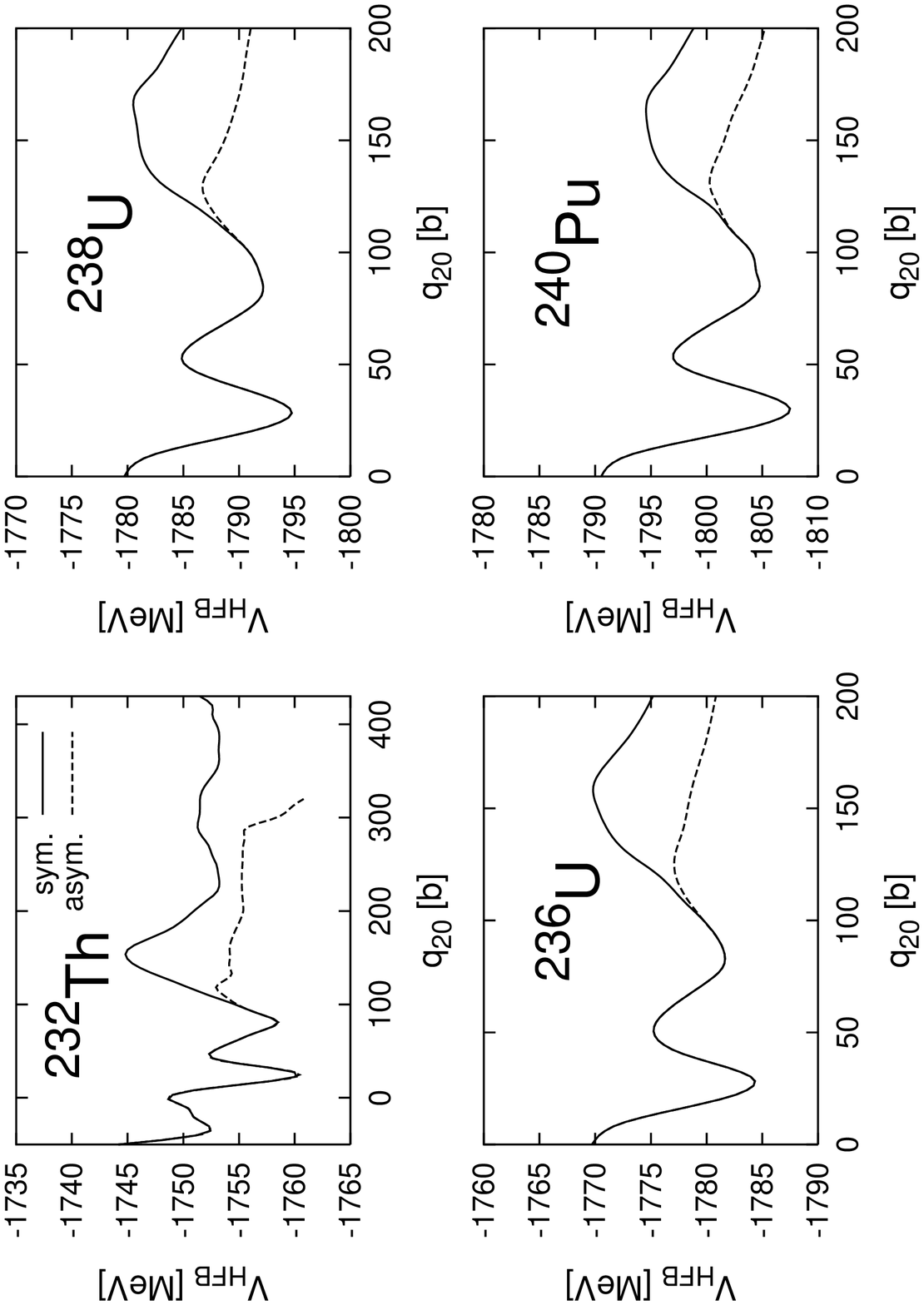}
\caption{ }
\end{figure}

\pagebreak[5]
\begin{figure}
\epsfxsize=160mm \epsfbox{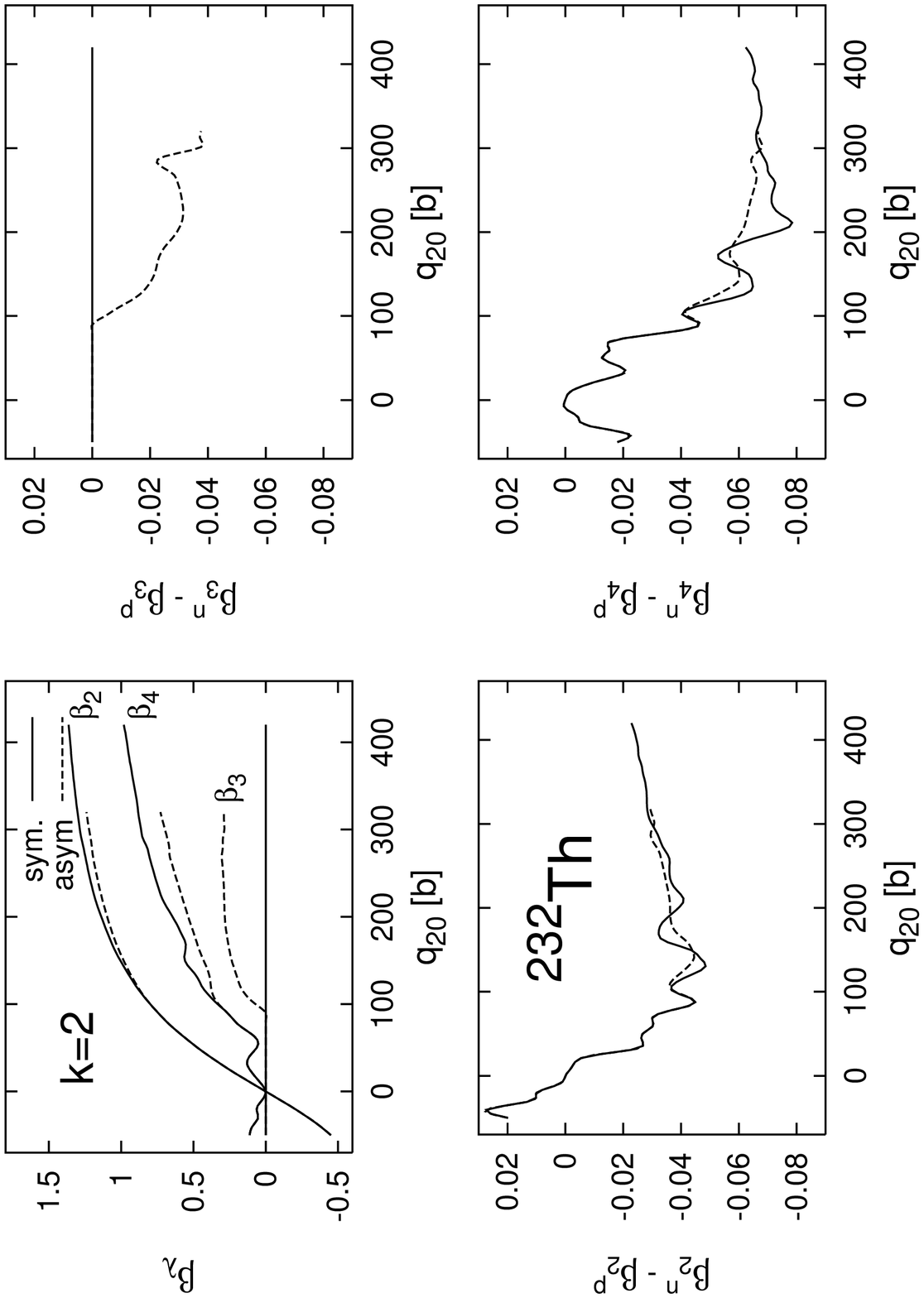}
\caption{ }
\end{figure}

\pagebreak[5]
\begin{figure}
\epsfxsize=160mm \epsfbox{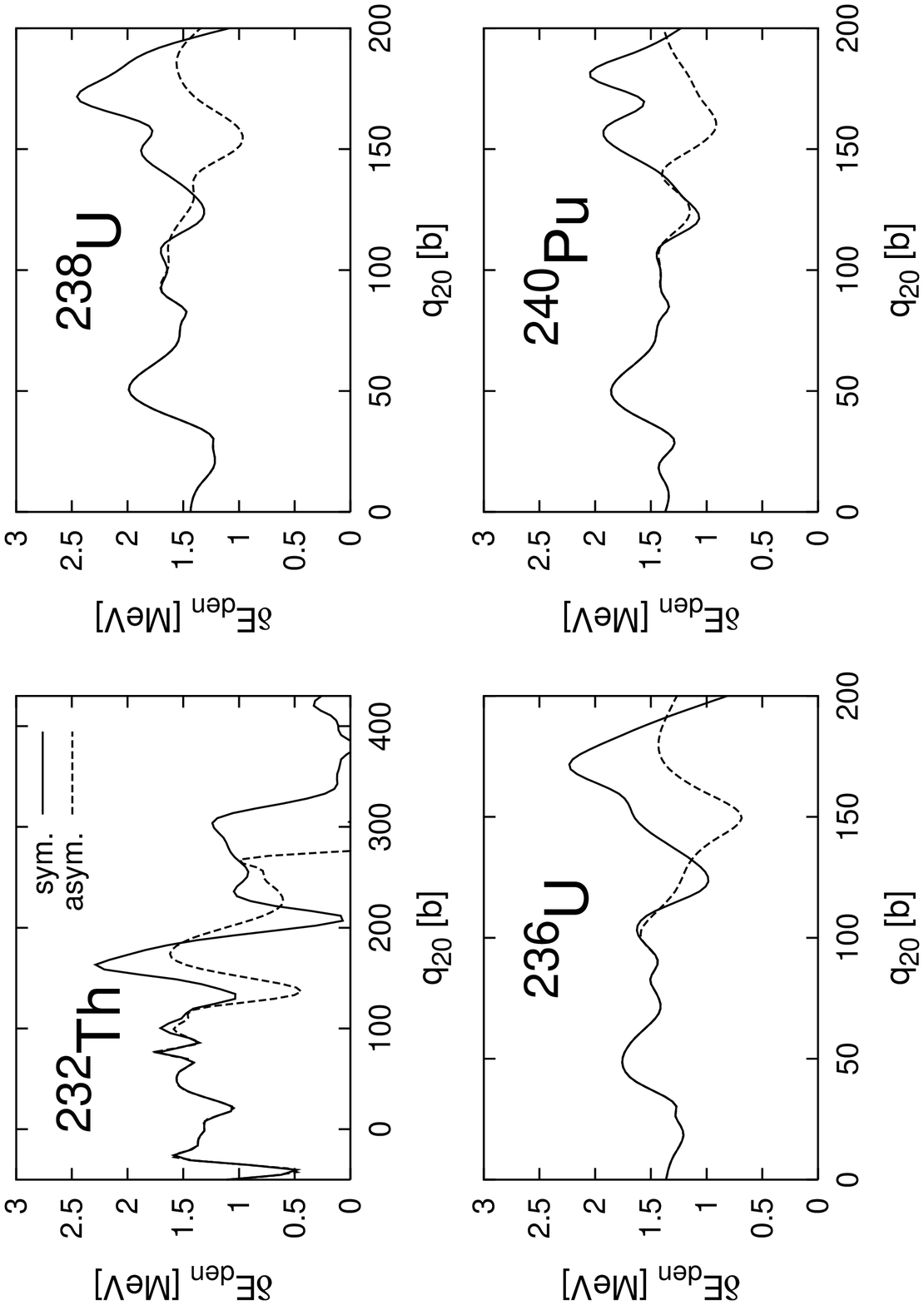}
\caption{ }
\end{figure}

\pagebreak[5]
\begin{figure}
\epsfxsize=160mm \epsfbox{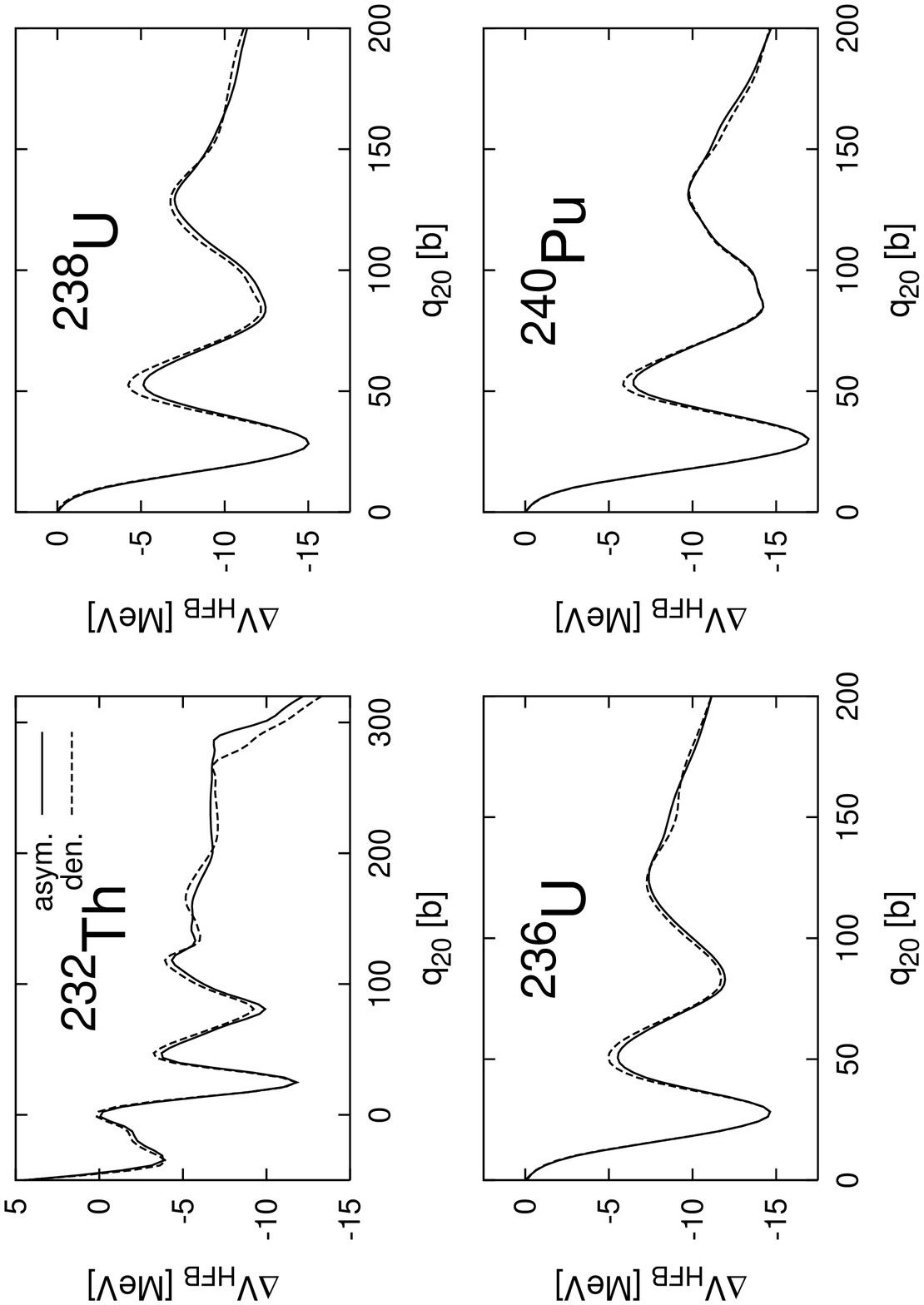}
\caption{ }
\end{figure}

\end{document}